\documentclass[12pt]{article} 
\usepackage{amsmath,amssymb}

\usepackage{color}

\usepackage[margin=1.in]{geometry}

\usepackage{graphicx}
\usepackage{{float}}
\usepackage{enumitem}

 \newcommand{\bea}{\begin{eqnarray}}
\newcommand{\eea}{\end{eqnarray}}
\newcommand{\be}{\begin{equation}}
\newcommand{\ee}{\end{equation}}
\newcommand{\ba}{\begin{align}}
\newcommand{\ea}{\end{align}}

\newcommand{\ZZ}{\mathbb{Z}} 

\newcommand{\Tr}{{\rm {Tr}}}

\newcommand{\Z}{\mathbb{Z}}

\begin{document}

\title{\bf{Monstrous Product CFTs
\\[2mm] in the Grand Canonical Ensemble\\[0.6 cm]}}

\author{Paul de Lange$^\heartsuit$, Alexander Maloney$^\spadesuit$, and Erik Verlinde$^\clubsuit$$^\diamondsuit$ }
\date{}

\maketitle

\begin{center}

$^\heartsuit$ {\it Department of Physics and Astronomy, University of Kentucky, Lexington, USA}\\

\vspace{0.3cm}

$^\spadesuit$ {\it  Department of Physics, McGill University, Montr\'eal, Canada } \\  

\vspace{0.3cm}

$^\clubsuit${\it Institute of Physics, University of Amsterdam, the Netherlands}

\vspace{0.3cm}

$^\diamondsuit${\it Delta-Institute for Theoretical Physics, Amsterdam, the Netherlands}

\end{center}
\vspace{1cm}

\begin{center}
{\bf Abstract}
\end{center}
We study symmetric products of the chiral `Monster' conformal field theory with $c=24$ in the grand canonical ensemble by introducing a complex parameter $\rho$
conjugate to the number of copies of the CFT.  The grand canonical partition function is given by the DMVV product formula in terms of the multiplicities of the seed CFT. It possesses an $O(2,2;\ZZ)$ symmetry that enhances the familiar $SL(2,\ZZ)$ modular invariance of the canonical ensemble and mixes the modular parameter $\tau$ with the parameter $\rho$. By exploiting this enhanced modular symmetry and the pole structure of the DMVV product formula we are able to extend the region of validity of Cardy's formula, and explain why it matches the semi-classical Bekenstein-Hawking formula for black holes all the way down to the AdS-scale. We prove that for large $c$ the spectrum contains a universal perturbative sector whose degeneracies obey Hagedorn growth.  The transition from Cardy to Hagedorn growth is found to be due to the formation of a Bose-Einstein condensate of ground state CFTs at low temperatures. The grand canonical partition function has an interesting phase structure, which in addition to the standard Hawking-Page transition between low and high temperature, exhibits a wall-crossing transition that exchanges the roles of $\tau$ and $\rho$. 

\newpage

\tableofcontents
\newpage

\section{Introduction}

Essentially every microscopic explanation of black hole entropy in string theory is based on Cardy's formula for the  
asymptotic density of states of a two dimensional conformal field theory.  The first of these calculations is the celebrated result of Strominger and Vafa, who reproduced the entropy of supersymmetric black holes by counting D-brane bound states using an effective 2D CFT \cite{Strominger:1996sh}.  The relationship between Cardy's formula and black hole entropy is particularly clear in the holographic correspondence between 3D Anti-de Sitter space (AdS$_3$) and 2D conformal field theory (CFT$_2$),  where the CFT density of states matches the Bekenstein-Hawking entropy of the $BTZ$ black hole \cite{Strominger:1997eq}.  

Cardy's formula for the asymptotic density of states with left- and right-moving dimension $(\Delta, {\bar \Delta})$ is \cite{Cardy:1986ie}
\be
\label{cardy}
S_{Cardy}
= 4 \pi \sqrt{ {c\over 24}\left(\Delta - {c\over 24}\right) } + 4 \pi \sqrt{ {c\over 24}\left({\bar \Delta} - {c\over 24}\right) }
\ee
Here $c$ is the central charge of the theory, which in the dual theory of AdS$_3$ gravity
is related to Newton's constant $G$ by \cite{Brown:1986nw}
\be\label{c}
c = {3 \ell_{AdS} \over 2 G}~.
\ee
States in the CFT with large $(\Delta, {\bar \Delta})$ are regarded as microstates of the BTZ black hole, 
whose horizon radius $r$ is given by
\be\label{delta}
\left(\Delta - {c\over 24}\right) = {r^2 \over 16 \ell_{AdS} G}~, 
\ee
where $\ell_{AdS}$ is the AdS radius.\footnote{For simplicity we have given here the formula for a non-rotating black hole with $\Delta = {\bar \Delta}$.  The matching works just as well in the rotating case.} 
With these identifications, Cardy's formula then matches precisely the Bekenstein-Hawking formula for the entropy of the BTZ black hole.

%
%

This impressive result hides an important puzzle, as Cardy's formula and the Bekenstein-Hawking formula have two different regimes of validity.    The Bekenstein-Hawking formula should be valid for semi-classical theories of AdS gravity, where Newton's constant is sent to zero while keeping the AdS radius fixed.  Moreover, it should hold when the area of the black hole is large in Planck units even if it is not necessarily large in AdS units.  In the CFT language, this happens when we are in the:
\be
\label{bhr}
{\rm \bf BH~Regime:} ~~~ c\to \infty\quad {\rm with}~{\ \Delta/c\ }~{\rm fixed}.
\ee
The Cardy formula, on the other hand, holds for any CFT with a fixed central charge $c$.  The only requirement is that the dimension $\Delta$ is large compared to the central charge $c$.  Hence the regime of validity of the Cardy formula is the
 \be
 \label{cardyr}
{\rm \bf Cardy~Regime:} ~~~ \Delta/ c \to \infty\quad {\rm with}~{\ c\ }~{\rm fixed}.
\ee
We are thus presented with a puzzle: why does Cardy's formula work at all in the BH regime? 


First, we note that in order to even define a semi-classical limit one must consider not a single CFT but rather a family of CFTs with a large central charge limit.  We will label our CFTs by a parameter $N$, with the property that $c\to\infty$ as $N\to\infty$.  In order for a dual gravity description to exist the family of CFTs must  obey a number of constraints at large $N$: for a discussion of these conditions see e.g. \cite{Heemskerk:2009pn, ElShowk:2011ag, Belin:2014fna}.   
One can then ask under what circumstances Cardy's formula will apply in the BH regime.
For example, the authors of \cite{Hartman:2014oaa} (see also \cite{Keller:2011xi, Maloney:2007ud}) showed that one must require that the density of perturbative states, i.e. those with dimension less than $\Delta - c/24$, does not grow too quickly. 
%

We will give a somewhat different perspective on this problem, which
is motivated by the following observation.  The original Cardy formula (\ref{cardy}), valid in the Cardy regime, follows from a  symmetry of the theory: modular invariance.    It is natural to ask whether there is a similar symmetry argument which guarantees the validity of the Cardy formula in the BH regime. 
We will answer this question in the affirmative for a certain special class of 2D conformal field theories.
Our argument will rely on a ``dual" version of modular invariance, which relates the behaviour of a family of CFTs at large $N$ to the behaviour at small $N$, just as standard modular invariance relates the low energy and high energy behaviours. %

We will focus in this paper on chiral CFTs, which possess only left-moving degrees of freedom.  In this case the right-moving sector is trivial, and only the first term in Cardy's formula (\ref{cardy}) is present. The dual theories of AdS gravity are somewhat more subtle, and include a gravitational Chern-Simons term whose coefficient is fixed by the constraint that the right-moving central charge to zero and modifies the Bekenstein-Hawking formula (see e.g. \cite{Kraus:2006wn, Li:2008dq, Maloney:2009ck} for a discussion of such theories).  However, our focus is not the gravitational interpretation of these theories; our goal is simply to find a class of theories for which the validity of Cardy's formula can be extended into the BH regime.

To this end, we will consider symmetric product CFTs, with central charge\\[-2mm]
$$
c= 24 N.
$$ 
In particular, we will consider the $N$-fold symmetric products of a special chiral `seed' CFT with central charge $c=24$.  There exist a number of chiral CFTs at $c=24$ \cite{Schellekens:1992db}, but for definiteness we will choose the so-callled ``Monster CFT," whose partition function is given by the familiar\footnote{We will use a convention which differs slightly from the usual one in the mathematics literature, where $J(q)$ has been shifted so that its constant term vanishes.} Klein invariant $J(q)$:\\[-1mm]
\begin{equation}
\label{jfunction}
J(q) =  \sum_{m\geq -1} c(m) q^m =  q^{-1} + 196884 q+....
\end{equation}
This choice is mostly (but not entirely) made in order to keep our formulas simple: many of our results can be straightforwardly generalized to any chiral CFT. The generalization to non-chiral CFTs is more subtle, however, and will be left to future work. 

Symmetric product CFTs are not expected to be dual to Einstein gravity. This can be seen by investigating the density of states at large $\Delta$, but with $\Delta <c/24$; such states are too light to form black holes, so can be used to characterize the perturbative spectrum of states in AdS.
The perturbative spectrum of a symmetric product CFT has Hagedorn behavior \cite{Keller:2011xi}, such as would be found in a semi-classical string theory with string tension proportional to the AdS scale. 
Nevertheless the spectrum of states for $\Delta> c/24$  will be governed by the Cardy formula, which for these CFTs has a range of validity that precisely corresponds to that of the semi-classical Bekenstein-Hawking formula for the BTZ black hole.

A central observation of this paper is that this result can be understood from a dual form of modular invariance as follows.
For each CFT in this family, it is convenient to package the spectrum into the partition function
\be
Z_N(\tau) \equiv \Tr~ q^{(\Delta-c/24)}
\qquad \qquad q=e^{2\pi i\tau}
\ee 
where the trace is taken over the Hilbert space of the $N^{th}$ conformal field theory:
$Z_N(\tau)$ is a standard canonical ensemble partition function, where ${\rm Im}~\tau$ plays the role of  the inverse temperature.  
Regarding $\tau$ as the modular parameter of a torus, $Z_N(\tau)$ is invariant under the usual $SL(2,\ZZ)$ transformations describing the large conformal transformations of the torus
$$
Z_N(\tau) = Z_N\left({a \tau + b\over c \tau+d}\right)
$$
The CFT spectrum is thus constrained by modular invariance, which leads to the usual derivation of Cardy's formula at fixed $N$.  To guarantee the validity of Cardy's formula in the BH regime we will employ a grand canonical ensemble, by introducing a second parameter $\rho$ whose imaginary part represents a chemical potential dual to the integer $N$.  The grand canonical partition function  is defined as 
\be\label{grand1}
{\cal Z}(\rho,\tau) =\sum_{N=0}^\infty p^{N+1} Z_N(q) \qquad\mbox{with}\qquad p=e^{2\pi i\rho},\,\,\,\,Z_0(q)=1
\ee
and will be our main object of study.

Our central observation is that this grand canonical partition function possesses a ``dual" modular symmetry, under which $\rho$ transforms under $SL(2,\Z)$.  Just as the $\tau \to -1/\tau$ symmetry relates high and low temperature, the transformation $\rho \to -1/\rho$ relates small and large values of $N$.  It is this dual version of modular invariance which guarantees the validity of Cardy's formula in the BH regime.  We will discover that this grand canonical partition function encodes many other interesting properties of the theory. 

The outline of the paper is as follows. We begin in section 2 by reviewing the derivation of Cardy's formula, its range of validity and its gravitational interpretation.  In section 3 we present the grand canonical partition function of the monster CFT and discuss its symmetries and pole structure. These results are then used in section 4 to derive the asymptotic growth properties of the micro-canonical degeneracies. We will show that for $\Delta > c/12$ the degeneracies obey Cardy growth, while for $\Delta < c/12$ they obey Hagedorn growth. We will demonstrate that in the second regime the ground states form a Bose-Einstein condensate, which leads to a physical explanation for the different growth properties. For $\Delta < c/24$ it is found that the degeneracies are universal and describe a perturbative gas of tensionless strings. In section 5 we compute the partition function of these universal perturbative states and use this result to present an exact semi-classical expansion of the canonical partition function (also known as a `Farey tail'). A key role in our analysis is played by the pole at $p=q$, whose residue precisely gives the perturbative states. In section 6 we again consider the symmetries of the grand canonical partition function and study their implications for the micro-canonical degeneracies. We find that the presence of the $p=q$ pole breaks the apparent $Z_2$ symmetry that exchanges $p=q$ at the micro-canonical level due to a phenomenon that is analogous to wall crossing. We explain in detail how the properties of the micro-canonical degeneracies can be made consistent with the full symmetry of the grand canonical partition function. Finally, in section 7 we present the conclusions and provide an outlook on possible future directions.   


 
\section{Review: Cardy's Formula and its Range of Validity}

We begin with a review of the derivation of Cardy's formula \cite{Cardy:1986ie}, which follows from the modular invariance of the 
CFT partition function\\[-1mm]
\be
Z(\tau) = \sum_\Delta D_c(\Delta) e^{2\pi i \tau (\Delta-c/24)} 
\ee
where $D_c(\Delta)$ denotes the number of states with conformal dimension $\Delta$.\footnote{As we are considering a chiral CFT, the dimension $\Delta\in \Z$ is quantized.}  We have added the label $c$ to indicate the number of states in a  fixed theory with central charge $c$. 
The modular transformation $\tau \to -1/\tau$ relates the behavior of the partition function at $\tau\to 0$ to that at $\tau\to i\infty$, where the partition function is dominated by the vacuum state with $\Delta = 0$.  Including the shift in the ground state energy due to the conformal anomaly, this gives the high temperature limit:\\[-1mm]
\be
\label{zapprox}
\lim_{\tau\to 0} Z(\tau) = \exp \left({2\pi i \over\tau}{c\over 24}\right) 
+\dots
\ee
where $\dots$ represents terms that are exponentially suppressed as $|\tau|\to 0$. 
The asymptotic behaviour of the degeneracies can be obtained by writing the density of states as an inverse Laplace transform of $Z(\tau)$, and inserting this high temperature estimate:
\be
\label{integral}
D_c(\Delta) = \int_{-i\infty}^{i\infty} \!\!d\tau \exp \left(-2\pi i \tau (\Delta-c/24)+{2\pi i \over\tau}{c\over 24}\right)
+\dots
\ee
At large $\Delta$ the integral has a saddle point at $\tau_o = i \sqrt{c/24 \over {\Delta}-c/24}$, where 
the saddle point approximation gives the familiar result 
\be
D_c(\Delta) = \exp \left\{4\pi \sqrt{{c\over 24}\left(\Delta-{c\over 24}\right)}\right\}+\dots
\ee 

We now come to the important point, which is the regime of validity of Cardy's formula.  
In a general CFT we expect the approximation (\ref{zapprox}) to fail once $|\tau|$ becomes of order one. 
This means that our saddle point approximation is valid only when the ratio\\[-1mm]
\be
\label{deltabd}
{\Delta  - c/24 \over c/ 24} >\!\!>\! 1
\ee
is large.

\subsection{Semi-classical gravity expectations}

We can compare this with our expectations for a holographic theory dual to AdS$_3$ gravity. 
In this case a state with large $\Delta$ is dual to a black hole with horizon radius $r$ given by equation (\ref{delta}).  So, using the Brown-Heanneaux formula (\ref{c}) for the central charge, this means that Cardy's formula applies only to black holes with \\[-1mm]
$$
{r \over \ell_{AdS}} >\!\!>\! 1~.
$$
So Cardy's formula reproduces the entropy of black holes whose horizon size is much larger than the AdS radius.
This is quite far from our semi-classical expectation, which is that the Bekenstein-Hawking formula should hold whenever the area of the black hole is large in Planck units:\\[-2mm]
$$
{r \over G} >\!\!>\! 1
$$
which would occur whenever
\be
{c \over 24} \left({\Delta  - {c\over 24} } \right)  >\!\!>\! 1
\ee
is large, rather than just the ratio (\ref{deltabd}).  

The above argument is still somewhat naive, however, as a black hole with $\Delta > c/24$ will have non-zero entropy but may still fail to dominate density of states -- its entropy must exceed that of whatever other states are present in the theory, such a gas of perturbative fields or strings in an AdS background.   So the precise location of the transition between a perturbative regime and a black-hole regime will depend on the nature of the semi-classical theory.  For example, if the perturbative spectrum is extremely sparse -- such as in the putative extremal CFTs which are conjectured to be dual to pure gravity \cite{Witten:2007kt} -- then the transition will occur at $\Delta=c/24$ in the large $c$ limit.  
For theories with a higher density of perturbative states, such as string theories, the transition will occur at a larger value of $\Delta$.
This can be made more precise by considering the gravitational computation of the canonical ensemble partition function.
In this computation, one expects a phase transition precisely at $\tau_o=i$, where two gravitational saddle points (thermal AdS and the BTZ black hole) interchange dominance.  From our previous formula for $\tau_o$, this occurs at $\Delta=c/12$.  Thus our general expectation is that a phase transition will occur somewhere in the ``enigmatic regime," where $\Delta$ lies between $c/24$ and $c/12$.  

In particular, we expect that the transition should occur at a value of $\Delta$ which is no larger than $c/12$.
Thus our expectation is that in any CFT with a semi-classical  gravity dual, the Cardy formula should apply to states with
\be\label{cond}
{c \over 24} \left({\Delta  - {c\over 24} } \right)  >\!\!>\! 1~~~~{\rm and}~~~~ \Delta>c/12~.
\ee
%
This leads to the central question: under what circumstances is Cardy's formula valid in the regime (\ref{cond}), and does its validity in this regime follow from a symmetry of the partition function?  

To answer this question, it will be crucial to consider not just a single CFT but rather a family of CFTs where the central charge can be varied to take a semi-classical limit.  
Indeed, it is only in the limit of large central charge that we can cleanly divide the spectrum into a perturbative regime, consisting of those states  which are too light to form a black hole, and genuine black hole states.  
We will therefore consider a family of CFTs labeled by an integer $N$ which parameterizes the number of degrees of freedom, with $c\to\infty$ as $N\to\infty$. 
We will moreover focus on a single very special family of CFTs, where the power of modular invariance be extended to a remarkable degree.





 \section{Product CFTs in the Grand Canonical Ensemble}

We will consider a special family of CFTs: 
chiral CFTs with central charge $c=24N$, which are obtained by taking symmetric products of a seed CFT with central charge $c=24$. Chiral CFTs with $c=24$ are rare \cite{Schellekens:1992db}, and their partition functions are almost completely fixed by modular invariance. 
A well known example is the Monster CFT \cite{Frenkel:1988xz} whose partition function is equal to the familiar Klein invariant $J(q)$, with q-expansion given in equation (\ref{jfunction}).  The Monster CFT is obtained by considering 24 chiral bosons on the Leech lattice and orbifolding by a $\mathbb{Z}_2$ symmetry, and played a central role in the proof of the Moonshine conjecture for the Monster group \cite{Borcherds:1992aa}. 
Most of our discussion easily generalizes to other chiral CFTs. 

\subsection{Symmetric products and the DMVV formula.}

The partition function of the $N$-fold symmetric product can be obtained from the seed CFT using the DMVV product formula \cite{Dijkgraaf:1996xw, Dijkgraaf:1998zd}. This formula gives the partition function in the grand canonical ensemble where one introduces a chemical potential for the integer $N$ in addition to the temperature.

The grand canonical partition function ${\cal Z}(p,q)$ is defined by 
\be
\label{Zdef}
{\cal Z}(p,q) = \sum_{N=0}^{\infty} p^{N+1} Z_N(q) = \sum_{N\geq 0\atop M\geq -N} D(N,M) p^{N+1} q^M ~.
\ee
Here  $Z_N(q)$ is the canonical partition function and
 the integer $M$ is the shifted conformal dimension $\Delta -c/24$.
As  shown in \cite{Dijkgraaf:1998zd}, ${\cal Z}(p,q)$ is given 
by the product formula 
\be
\label{dmvv}
{\cal Z}(p,q) = p \prod_{n\geq 1\atop m\geq -1} (1-p^nq^m)^{-c(nm)}
\ee
where $c(n)$ is the number of states in the seed CFT with conformal dimension $\Delta=n+1$.
This formula makes clear that the states labeled by $N$ and $M$ can be constructed out of a ``free gas" of states labeled by $n$ and $m$ and with multiplicity $c(mn)$. We have included an extra factor of $p$ in the definition (\ref{Zdef}), which also appears in (\ref{dmvv}), for future convenience. 

Our plan is as follows: first we will derive some important special properties of the grand canonical partition function ${\cal Z}(p,q)$ that may not be immediately obvious from its definition. These special properties deal with its symmetries and its pole structure. Specifically, we will show that the grand canonical partition function is automorphic with respect to 
an 
	$O(2,2;\mathbb{Z})$
 duality group, which contains two $SL(2,\mathbb{Z})$ subfactors that act on $\tau$ and $\rho$ in the familiar way.  
The duality group has in addition a $\mathbb{Z}_2$ factor that exchanges $\tau$ with $\rho$, or equivalently\\[-2mm]
$$
  p\ \leftrightarrow\  q
$$
This is a somewhat surprising and mysterious symmetry, since it exchanges the inverse temperature with the chemical potential. 
One of the main goals of this paper is to study the consequences of this exchange symmetry and examine its possible physical interpretation.  

The partition function ${\cal Z}(p,q)$ has a simple pole at $\rho=\tau$,  or equivalently at $p=q$. This `primary' pole is invariant under the exchange symmetry, but the other elements of the $O(2,2;\Z)$ symmetry map it to a different location. Hence the grand canonical partition function has poles at all its $SL(2,\mathbb{Z})$ images $\rho=\gamma({\tau})$, or equivalently at $p=q_\gamma$ where
\be
\label{poles}
q_\gamma = e^{2\pi i  \gamma(\tau)} 
\qquad\mbox{with}\qquad
\gamma(\tau) = {a\tau+b\over c\tau+d}~.
\ee In the following sections we will use this property of ${\cal Z}(p,q)$ to constrain the asymptotic behavior of the degeneracies $D(N,M)$ and derive the expected properties for a conformal field theory with a gravity dual.

Specifically, we will prove the following statements about the degeneracies $D(N,M)$:
We will demonstrate the following asymptotic properties, valid when $N,M >\!\!>1$:
\begin{itemize}
\item For $M>N $ the degeneracy $D(N,M)$ obeys the Cardy formula: \ 
$$
D(N,M) \sim \exp 4\pi\sqrt{NM} \qquad\qquad\mbox{for}\quad M > N >\!\!>1.
$$
\item  For $N>M$ the degeneracy $D(N,M)$ exhibits Hagedorn growth:\  
$$
D(N,M) \sim \exp 2\pi (N+M) \qquad\qquad\mbox{for}\quad N > M >\!\!>1~.
$$
\end{itemize}
We will moreover prove the following exact statements for all $N,M$:
\begin{itemize}
\item  For $M<0$ the degeneracy $D(N,M)$ is universal and depends only on $\Delta = N+M$:
$$
D(N,M) = D_\infty(N+M) \qquad\qquad\mbox{for all}\quad M<0.
$$	
\item  The degeneracy $D(N,M)$ is always smaller than the universal density $D_\infty(N+M)$ by an amount that is precisely equal to $D(M-1,N+1)$ 
$$
D(N,M)  = D_\infty(N+M) - D(M-1,N+1) \qquad\qquad\mbox{for all $M$ and $N$}.
$$	  
\end{itemize}

The first result is precisely Cardy's formula, but with the extended range of validity expected of a CFT with a gravity dual. The third property, which states that the perturbative part of the spectrum (describing states below the black hole threshold) is independent of the central charge, is consistent with such an interpretation. The Hagedorn behavior for the states with $M<N$  is not expected to be generic for all CFTs, since it implies that the perturbative spectrum is not described by a conventional particle or field theory. Rather, it appears that the symmetric product CFTs at the orbifold point are described by a low tension string theory, whose string scale is set by the AdS radius.  Note that the combination of the second and third property implies that also the universal asymptotic spectrum obeys Hagedorn scaling.
The physical significance of this fact will be further discussed below.

\subsection{Symmetries and pole structure of ${\cal Z}(p,q)$.}

The key ingredients in the derivation of these properties are the above mentioned symmetries and pole structure of the grand canonical partition function ${\cal Z}(p,q)$. We will start with a proof of these special properties. 
To begin, we emphasize that the partition function ${\cal Z}(p,q)$ is defined for values of $p$ and $q$ with $|p|$ and $|q|$ both smaller than one. In other words, $\rho$ and $\tau$ live in the upper half complex-plane. 

One striking feature of the product formula (\ref{dmvv}) for the grand canonical partition function is that it is almost symmetric in $p$ and $q$. 
Note that the range of the integers $n$ and $m$ is slightly different, however:  $m$ is taken from minus one to infinity, while $n$ is strictly positive. Since $c(nm)$ is non-zero only for $nm\geq -1$, we conclude that the asymmetry in $p$ and $q$ is entirely due to the single factor with $m=-1$ and $n=1$.  
Since $c(-1) =1$ we can explicitly remove this factor, by writing
\be
{\cal Z}(p,q) ={\cal Z}_0(p,q) {\cal Z}_>(p,q)
\ee
where ${\cal Z}_>(p,q)$ represents the contribution of only the excited states. 
\be
\label{Zgreater1}
{\cal Z}_>(p,q) = \prod_{n\geq 1\atop m\geq 1} (1-p^nq^m)^{-c(nm)}~.
\ee
The prefactor ${\cal Z}_0(p, q)$ describes the gas of ground states of the seed CFT, and is given by the factor with $m=1$ and $n=-1$, along with the overall factor $p$ introduced in the definition of ${\cal Z}(p, q)$. 
Indeed, this factor was introduced so that ${\cal Z}_0(p,q)$ can be written as 
\be
\label{Zzero}
{\cal Z}_0(p, q)= {1\over p^{-1}-\,q^{-1}} 
\ee
which is anti-symmetric in $p$ and $q$. 
The partition function ${\cal Z}_>(p,q)$ of the excited states is manifestly symmetric in $p$ and $q$, so the full grand canonical partition function is anti-symmetric in $p$ and $q$:
\begin{equation}
{\cal Z}(p,q) 
= - {\cal Z}(q,p) \label{antisym}
\end{equation}

The individual canonical ensemble partition functions $Z_N(q)$ are invariant under the 
modular $SL(2,\mathbb{Z})$ transformations that act on $q=e^{2\pi i \tau}$  in the usual manner.
This implies that the grand canonical partition function also obeys the same symmetry. It is important to note that the $\mathbb{Z}_2$ that exchanges $p$ and $q$ does not commute with this symmetry. Instead, it exchanges the $SL(2,\mathbb{Z})$ that acts on $\tau$ with another $SL(2,\mathbb{Z})$ group action on $\rho$. The full symmetry group of ${\cal Z}(p,q)$ is therefore
$$
O(2,2;\mathbb{Z})=\Bigl(SL(2,\mathbb{Z})\times SL(2,\mathbb{Z})\Bigr)\ltimes \mathbb{Z}_2 
$$
Indeed, as was shown in \cite{Borcherds1995}, the function (\ref{dmvv}) is an automorphic form on this group. This symmetry group is equal to that of a complex scalar CFT on a two-torus $\mathbb{T}^2$. Indeed, it is possible to 
relate the product formula to the integrated one loop free energy of such a CFT \cite{Harvey:1995fq}. For the present 
purpose this connection is not immediately relevant, although it may play a role in understanding the 
interpretation of the parameter $\rho$ in a dual gravitational description. We will come back to this point in 
the discussion.

The degeneracies $D(N,M)$ are obtained from ${\cal Z}(p,q)$ via an inverse Laplace transform 
\be
\label{contour}
D(N,M) =\oint {dp\over 2\pi i}\, \oint {dq\over 2\pi i} 
 \, p^{-N-2} \,q^{-M-1} {\cal Z}(p,q)~.
\ee
The value of this integral depends on the choice of integration contour for the $\rho$ and $\tau$ variables. 
Since the integrand is meromorphic in both variables, the integrals can in principle be evaluated by Cauchy's theorem in terms of the residues at the various poles. As we will now discuss, the duality symmetry serves as a useful tool to determine these poles.

Naively, the product expression (\ref{dmvv}) has an infinite number of poles, namely at all values for which $p^nq^m=1$. However, none of these poles lie in the right domain (the upper-half planes of $\tau$ and $\rho$), except for the case $n=1$ and $m=-1$. Hence the partition function has a simple pole at 
$p=q$. Clearly, this pole is invariant under the $\mathbb{Z}_2$ action that interchanges $p$ and $q$. But the modular $SL(2,Z)$ symmetry that acts on $p$ or $q$ maps it to different poles that are located at $p=q_\gamma$. Thus 
when we restrict $\rho$ and $\tau$ to the fundamental domain, there is only one simple pole. In 
addition, the partition function vanishes when either $\rho$ or $\tau$ are taken to $i\infty$. These facts are 
sufficient to determine the grand canonical partition function uniquely.

The Klein invariant $J(q)$ has only a simple pole at $q=0$ (or at $\tau=i\infty$) and hence it maps the modular domain exactly once onto the entire complex plane. 
The fact that the only poles of ${\cal Z} (p,q)$ are located at $p=q_\gamma$ and its only zeros are at $i\infty$ 
tells us that it must be given (up to a multiplicative factor that is fixed by the seed partition function) by 
\be\label{borcherds}
{\cal Z}(p,q) = {1\over J(p)-J(q)} 
\ee
Since the coefficients $c(n)$ are also taken from the $J$-invariant itself 
we recover a famous result due to Borcherds: 
\be
\label{borcherdsformula}
 \prod_{n\geq 1\atop m\geq 1}(1-p^nq^m)^{-c(nm)}   ={p^{-1} - q^{-1} \over J(p)-J(q)} 
\ee
On the left hand side we recognize the excited state contribution ${\cal Z}_>(p,q)$ given in (\ref{Zgreater1}), while the right hand side is precisely equal to 
the ratio of ${\cal Z}(p,q)$ with the ground state contribution ${\cal Z}_0(p,q)$ defined in (\ref{Zzero}).  The Borcherds formula  (\ref{borcherdsformula}) will play an important role in our following discussion, where it will interpreted as a modular-invariant generalization of the familiar partition function for a Bose-Einstein condensate.

As a preparation for this discussion, let us note that equation (\ref{borcherdsformula}) in particular implies that the expression (\ref{Zgreater1}) for ${\cal Z}_>(p,q)$  no longer has a pole at $p=q$, and thus has a well defined limit for $q\to p$. The limit value is easily computed using l'H\^opital's rule: 
\be
\label{comment}
 \lim_{q\to p} {\cal Z}_>(p,q) =  \lim_{q\to p} {p^{-1} - q^{-1} \over J(p)-J(q)} = -{1\over p^2 J'(p)}= \prod_{n\geq 1\atop m\geq 1}(1-p^{n+m})^{-c(nm)} \\[2mm]
\ee
The physical significance and interpretation of this formula will be further clarified below. 
\newpage
\section{Cardy behavior versus Hagedorn growth}

In this section we will present a detailed analysis of the microscopic degeneracies $D(N,M)$ using the pole structure and symmetry properties of the grand canonical partition function. For this purpose we will find it convenient to separate the excited states from the contribution of the ground states. 
We will first prove that, excluding the contribution of the ground states, the excited states exhibit Cardy type growth for all large values of $N$ and $M$. Subsequently we will study the effect of including the ground states. This will modify the asymptotic behavior only for states with $M<N$, and lead to a Hagedorn spectrum in that regime. 

\subsection{General proof of Cardy formula for excited states}
Let us first consider the partition function ${\cal Z}_>(p,q)$ for the excited states, and introduce degeneracies ${\cal D}(N,M)$ via its series expansion in powers of $p$ and $q$ 
\be
\label{Zgreater}
{\cal Z}_>(p,q) =\sum_{N\geq 0,M\geq 0}{\cal D}(N,M) p^{N}q^{M}.
\ee
Since ${\cal Z}_>(p,q)$ is manifestly symmetric in $p$ and $q$, the excited state degeneracies ${\cal D}(N,M)$ obey
$$
{\cal D}(M,N)= {\cal D}(N,M)~.
$$
This is a strong and important result! In the usual derivation we expect that these degeneracies obey the Cardy formula in the regime where $M>N$, with $M$ and $N$ both large. The reason is that in this regime the exclusion of the ground state presumably does not severely change the asymptotic growth in this regime. The symmetry between $M$ and $N$ now suggests that the Cardy formula should also apply in the opposite regime, with $M<N$. Indeed, we will now present a proof of the fact that ${\cal D}(N,M)$ exhibits Cardy growth for large, generic values of $M$ and $N$. 

We start again from a contour integral representation analogous to (\ref{contour}):
\be
\mathcal{D}(N,M) = \oint \frac{dp} {2\pi i}\, \oint \frac{dq}{2\pi i}  
 \, p^{-N-1} \,q^{-M-1}\, {\cal Z}_>(p,q)
\ee
The key observation is that, due to the fact that we have removed the ground states, the poles occur at 
$$
p= q_\gamma \qquad\mbox{with}\qquad \gamma \neq 1
$$
In particular, we have a pole at $\rho = -1/\tau$. Performing the contour integral over $p$ and taking the residue at this pole we obtain
$$
{\cal D}(N,M) = \int {dq \over 2\pi i} 
{{}\  \widetilde{q}^{-N-1} \,q^{-M-1}} \ \, {\cal Z}_>(\widetilde{q} ,q)
$$
with
$$
\widetilde{q} = \exp\left(-{2\pi i\over\tau}\right)~.
$$
We have obtained exactly the same integral expression as equation (\ref{integral}) for the conventional derivation of the Cardy formula, but this time we did not have to make any assumption about the value of $q$! We can now follow the same steps as before, and evaluate the integral using the saddle point approximation. The factor containing the residue of ${\cal Z}(p,q)$ at $p=\widetilde{q}$ becomes unimportant for the evaluation of the saddle point if we take $N$ and $M$ both to be large. One easily verifies that the  saddle point value is again reached for $\tau^2 = -M/N$, leading to the familiar Cardy result
\be
{\cal D}(N,M) = \exp 4\pi\sqrt{NM}~.
\ee
This now holds for all large values of $N$ and $M$. We thus reach the important conclusion that by including only the excited states of the symmetric product CFT, we always have a Cardy growth. This Cardy growth results from taking the pole at $p=\tilde{q}$. Of course, one should verify that the other poles at $p=q_\gamma$ with $\gamma\neq 1$ and different from $\gamma: \tau \to -1/\tau$ only give sub-leading contributions. This turns indeed out to be the case, as is demonstrated in the appendix of \cite{Dijkgraaf:1996it}. We will also return to this point below.

\subsection{Including the ground states and Hagedorn growth}

The next step is to include the ground states. As we now show, this will change the asymptotic behavior only for $M<N$ into a Hagedorn density of states. To include the ground states we have to reinsert the factor ${\cal Z}_0(p,q)$ into the partition function. The full degeneracies $D(N,M)$ can be simply expressed in terms of the ${\cal D}(N,M) $ by evaluating the contour integral
\be
\label{DNM-contour}
D(N,M) = \oint {dp\over 2\pi i}\, \oint {dq\over 2\pi i} 
 \, p^{-N-2} \,q^{-M-1}\, {\cal Z}_0(p,q) {\cal Z}_>(p,q)
\ee
using the expansion of the ground state partition function
\be
\label{zerosum}
{\cal Z}_0(p,q) = \sum_{K=0}^\infty p^{K+1} q^{-K}  \qquad \mbox{for} \qquad |p|<|q|~. 
\ee
The extra factor $p$ cancels out, and 
we get 
\be
\label{DNM-sum}
D(N,M) = \sum_{K=0}^{N} {\cal D}(N-K, M+K) ~.
\ee
This gives the microscopic degeneracies for all values of $M$ and $N$ in terms of the excited state degeneracies ${\cal D}(N,M)$. For the latter we already established that they exhibit universal Cardy growth. We will use this fact to determine the asymptotic growth of the degeneracies $D(N,M)$. 
We will further show that the inclusion of the ground state leads to a universal spectrum of states with low conformal dimension.

We can now distinguish a number of cases. First we look at negative values of $M$. In this case the sum of $K$ in (\ref{DNM-sum}) starts at $-M$ (which is a positive number) and runs to $N$. 
In this case one finds that after summing over $K$ the result for $D(N,M)$ only depends on the conformal weight $\Delta =N+M$,
\begin{equation}
\label{negativeM-sumrule}
D(N,M)= \sum_{K=-M}^{N} {\cal D}(N-K,M+ K) 
= D_\infty(N+M) 
\qquad \qquad \mbox{for} \quad M<0~.
\end{equation}
Here, $D_{\infty}(\Delta)$ represents the expected universal spectrum! It is given by the sum of all excited degeneracies with the same total conformal dimension. 
\be
\label{Dinfty}
D_\infty(\Delta) \equiv \sum_{N,M\atop N+M=\Delta} {\cal D}(N,M)~.
\ee
The relation (\ref{negativeM-sumrule}) turns out to be a special case of a more general sum rule for the degeneracies $D(N,M)$, that we will be discussed further below. 

Physically, what happens is that the inclusion of the ground states allows one to lower the amount of central charge in the excited states. This changes the growth properties, since now we can optimize the central charge for a given value of the conformal dimension $\Delta =N+M$. 

To estimate the growth of these degeneracies it is convenient to approximate the sum over $K$ by an integral, and use the Cardy result for the excited degeneracies. In this we find that the universal spectrum of perturbative states $D_\infty(\Delta)$ has a Hagedorn growth, namely
\be
D_\infty(\Delta) \sim \int dK \exp 4\pi \sqrt{K(\Delta-K)} \sim \exp 2\pi \Delta~.
\ee
Here in the last step we inserted the saddle point value for $K$, which occurs at $K =\Delta/2$.  

Let us now discuss the other values of $M$. 
When $M<N$ we can use this same reasoning to show that the degeneracies $D(N,M)$ exhibit Hagedorn growth, since in this regime the sum (\ref{DNM-sum}) over $K$ reaches the saddle point value at $K= (N-M)/2$.   As soon as $M>N$ the saddle point value for $K$ can no longer be reached. In this case the term with the maximum growth is the one for $K=0$, hence we stay in the Cardy regime. We thus reach the important conclusion that the Hagedorn behavior terminates exactly when $M=N$, which is the critical value where also the Hawking-Page transition occurs. Hence, the Hagedorn temperature and Hawking-Page temperature are the same. 

\subsection{Bose-Einstein condensation of the ground states}

The fact that for $M<N$ the saddle-point value for $K$ is non-zero, implies that below the Hawking-Page transition, a finite fraction of the CFTs in the $N$-fold symmetric product are in their ground state.
This means that the transition from Cardy- to Hagedorn-growth can be interpreted as being due to Bose-Einstein condensation of the ground states. Physically what happens is that the condensate screens the central charge and in this way is responsible for the universality of the perturbative spectrum. 

To demonstrate that the appearance of the Bose-Einstein condensate, let us again separate the system in a ground state and excited state contribution and evaluate the expectation value of $N$ for these respective contribution. From the expansion (\ref{Zdef}) of ${\cal Z}$ and ${\cal Z}_0$ we find
$$
\langle N+1\rangle_{tot}  = p{\partial\over\partial p}\log {\cal Z}(p,q)= p{\partial\over\partial p}\log {\cal Z}_0 (p,q)+p{\partial\over\partial p}\log {\cal Z}_>(p,q)= \langle N+1\rangle_0 + \langle N\rangle_> 
$$ 
Explicit evaluation using the results (\ref{Zzero}) and (\ref{borcherds}) gives the following expressions for the ground state contribution and total contribution to the central charge. 
\be
\label{BEC-formulas}
\langle N+1\rangle_0 = {1\over 1-pq^{-1}} \qquad\mbox{and}\qquad \langle N+1\rangle_{tot} = - {pJ'(p)\over J(p)-J(q)}~.
\ee
We thus find that the fraction of the ground states compared to the total is given by
\be
\label{BE-fraction}
{\langle N+1\rangle_0\  \over \langle N+1\rangle_{tot}} = - {J(p) - J(q)\over p^{-1}-q^{-1}} {1\over p^2 J'(p)} ~.
\ee
Here we recognize the same factors that also appear in the grand canonical partition function ${\cal Z}_>(p,q)$ of the excited states. In fact, as we noted in equation (\ref{comment}), the factor involving $p^2 J'(p)$ can also be obtained as a limit of ${\cal Z}_>(p,q)$ for $p\to q$. In this way we find that the fraction (\ref{BE-fraction}) can be expressed as 
$$
{\langle N+1\rangle_0\  \over \langle N+1\rangle_{tot}} = {{\cal Z}_> (p,p)\over {\cal Z}_>(p,q)}=\prod_{n\geq 1\atop m\geq 1} \left({1-p^nq^m\over 1-p^{n+m}}\right )^{c(nm)}~.
$$
where in the last step we used Borcherds formula. These expressions make clear that, in the limit $q\to p$,  the fraction of central charge that is contained in the ground states compared to the total central charge goes to one.\footnote{In the physical regime one imposes $|p|\!<\!|q|\!<\!1$, hence strictly speaking the limit $q\!\to\! p$ can not be~reached.} 
\\[-12mm]

\begin{figure}[H]
		\includegraphics[width=14cm]{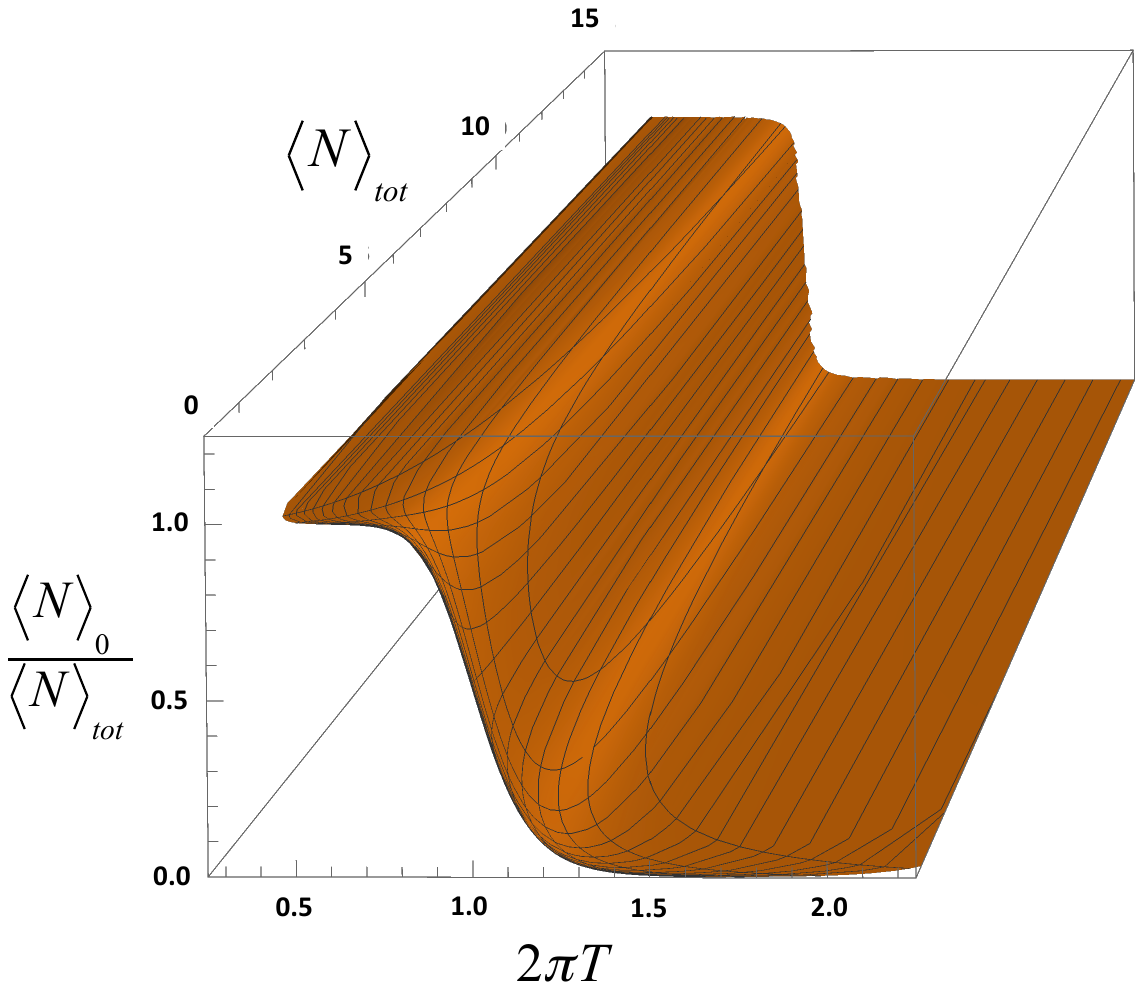}\vspace{-12mm}
		\centering
		\caption{This plot shows the ratio  $\langle N\rangle _0/\langle N\rangle_{tot}$ 
		as a function of the temperature $T$ for various values of $\langle N\rangle_{tot}$. The ratio $\langle N\rangle _0/\langle N\rangle_{tot}$ is very close to one below the transition temperature $T=1/2\pi$, indicating the existence of a Bose-Einstein condensate. 
		}
	\end{figure} 
 It can be verified numerically that the fraction of $\langle N\rangle _0/\langle N\rangle_{tot}$ abruptly jumps to one at the transition temperature $T=1/2\pi$, as illustrated in Figure 1. Here the temperature is defined in terms of $
 \tau$ via ${\rm Im\tau} = (2\pi T)^{-1} $. 
  For the numerical calculation we used the expression (\ref{BEC-formulas}), with $\rho$ and $\tau$ purely imaginary, and removed the shift by $1$ in both equations.   The fact the ratio $\langle N\rangle _0/\langle N\rangle_{tot}$ approaches the value one is a clear indication that a Bose-Einstein condensate of the ground states has formed in the low temperature regime.  

\bigskip

\section{The Perturbative Spectrum and an Exact Farey Tail}

In this section we will give an explicit formula for the partition function for the universal spectrum of perturbative states defined by 
\begin{equation}
\label{pertexp}
	{\cal Z}_{\infty}(q) \equiv \, \sum_\Delta D_\infty(\Delta) q^\Delta~.
\end{equation}
We will derive two different expressions for ${\cal Z}_{\infty}(q)$, by either using the DMVV product formula (\ref{dmvv}), or by using its relation to the $J$-function through Borcherds formula (\ref{borcherds}).  The key observation will be that ${\cal Z}_{\infty}(q)$ represents the residue at $p=q$ in the full grand canonical partition function ${\cal Z}(p,q)$. This fact is subsequently used to derive an exact Farey tail expansion \cite{Dijkgraaf:2000fq} (or Rademacher sum) for the canonical partition functions $Z_N(q)$ in terms of the first $N$ terms of  ${\cal Z}_{\infty}(q)$.

\subsection{Partition sum for the perturbative states} 

As explained above, the appearance of the universal perturbative spectrum  is closely related to the inclusion of the ground states. In fact, $D_\infty(\Delta)$, with $\Delta =N+M$, is the contribution in the contour integral (\ref{DNM-contour}) of the pole at $p=q$:
\begin{equation}
	D_{\infty} (\Delta) =-\oint {dq\over 2\pi i}  \,  q^{-(\Delta+1)} {\cal Z}_>(q,q) ~.
\end{equation}
This can be seen by inserting equation (\ref{Zgreater}) and performing the contour integral to directly recover the definition (\ref{Dinfty}) of $D_\infty(\Delta)$.  Thus the partition function ${\cal Z}_{\infty}(q)$ for the universal perturbative states can be written as 
\be
{\cal Z}_{\infty}(q) 
= \lim_{p\to q} {\cal Z}_>(p,q)
\ee
which can be evaluated by inserting the product formula (\ref{Zgreater1}) for ${\cal Z}_>(p,q)$ or by using the result (\ref{comment}) with $p$ and $q$ reversed.  This leads to a product formula for ${\cal Z}_{\infty}(q)$ which can be expressed in terms of the expansion coefficients $c(n)$ of the $J(q)$:
\be
\label{product}
{\cal Z}_{\infty}(q) =\prod_{\delta=1}^\infty (1-q^\delta)^{-d(\delta)}
\ee
where the multiplicities $d(\delta)$ are given by the sum of all microscopic multiplicities with fixed conformal dimension $\delta$
\be
\mathop{d(\delta) = \sum_{n+m=\delta}}_{\hspace{38pt}m\geq -1,\,n\geq 1} c(nm) ~.
\ee
This expression for $d(\delta)$ is similar to the one we found for $D_\infty(\Delta)$, except now defined in terms of $c(nm)$ instead of ${\cal D}(N,M)$. Since $c(nm)$ obeys the Cardy formula, we can use the same argument as before to show that $d(\delta)$ exhibits Hagedorn behavior:
$$
\mathop{d(\delta)}\ \sim\  \exp 2\pi \delta
$$ 
In other words, the perturbative spectrum behaves like that of a free gas of physical objects obeying Hagedorn growth. It is natural to interpret these objects as excited tensionless strings.

A second useful expression for the partition function ${\cal Z}_\infty(q)$ is given in terms of (the derivative of) the $J$-function 
 \be
\label{Zinfty3}
{\cal Z}_{\infty}(q)  
= -{1\over q^2J'(q)}
\ee
where we again made use of equation (\ref{comment}). 
This expression makes clear that the generating function ${\cal Z}_{\infty}(q)$ is a modular form of weight $-2$.  The product formula (\ref{product}), however, more clearly reveals the underlying physical interpretation as a gas of tensionless strings.  

\subsection{An exact Farey tail expansion}\label{Farey}

As explained above, the perturbative states represent the contribution of the residue at the primary pole at $p=q$. The full partition function has poles at all $SL(2,\mathbb{Z})$ images $p=q_\gamma$, which due to the full $O(2,2;\mathbb{Z})$ symmetry must have residues that are directly related to the partition function of the perturbative states. Thus the following identity must hold
\begin{equation}
\label{sumoverpoles}
{\cal Z}(p,q) = \sum_{\gamma} {{\cal Z}_\infty (q_\gamma)
\over p^{-1} - q_\gamma^{-1}}~.
\end{equation}
This expression exhibits all the poles and residues and is manifestly $O(2,2;\mathbb{Z})$ invariant. 
This allows us to derive an exact `Farey tail' formula \cite{Dijkgraaf:2000fq} for the canonical partition function $Z_N(q)$ by using its representation as as a contour integral of the grand canonical partition function. 
By performing the contour integration over $p$ and by summing over all residues we obtain an exact expression for the canonical partition function $Z_N(q)$ as a sum over all $SL(2,\mathbb{Z})$ images of the perturbative spectrum: 
\be
\label{fareytail}
Z_N(q) = \oint\! {dp\over 2\pi i} \,p^{-N-2} \mathcal{Z}(p,q) = \sum_\gamma q^{-N}_\gamma{\cal Z}_\infty (q_\gamma)~.
\ee
The sum clearly produces a modular invariant result. It is important to point out that the contour integral must be defined so that it indeed collects all the residues of poles while deforming the contour to infinity. This implies that we have to choose $|p|<|q|$.  
We can rewrite this result more explicitly in terms of the
perturbative degeneracies as
$$
Z_N(q) =
\sum_{\gamma ,\Delta }  D_\infty (\Delta)  q_\gamma^{\Delta-N}~.
$$
The fact that there are no modular functions without poles allows us to truncate the expressions in the sum to only the polar contributions. This means we only write the $q$-expansion up to the term with $q^{-1}$, and hence we only need to sum over a finite range of conformal dimensions  
$$
0\leq \Delta <N
$$
which is precisely the regime for which there are only perturbative states. Alternatively, we can insert the formulas (\ref{product}) and (\ref{Zinfty3}) that we obtained for the perturbative partition function to express the full canonical partition as 
\be
Z_N(q) = -\sum_\gamma \, \left.{q_\gamma^{-N-2}\over J'(q_\gamma)}\right|_{polar} \, =\, \sum_\gamma \left. {q_\gamma^{-N}\over  \ {\prod_{\delta} \,(1-q_\gamma^\delta)^{d(\delta)}}}\right|_{polar}~.
\ee
Here we again used the fact that we can restrict to polar terms. 
 It follows from these expressions that the full canonical partition function is uniquely fixed, and can be reconstructed from just the polar terms of the partition function, which contains only the perturbative part of the spectrum. 

\bigskip

\section{The $p\leftrightarrow q $ Exchange Symmetry and Wall Crossing}


 In this section we will focus on the role of the $\mathbb{Z}_2$ subgroup that interchanges $p$ and $q$ in the grand canonical partition function. This apparent $\mathbb{Z}_2$ symmetry of ${\cal Z}(p,q)$ naively suggests that all our previous results should also hold in an ensemble where we exchange the rescaled central charge $N\!= \!c/24$ for the shifted conformal dimension $\Delta- c/24\!=\!M$. The aim of this section is to examine this suggested symmetry in more detail.

 We will find that this symmetry is actually broken at the micro-canonical level by the contour prescription used to extract the microscopic degeneracies. The reason is that the pole at $p=q$ is intially situated at one side of the contour, but after the $\mathbb{Z}_2$ exchange operation the location of the contour is changed so that it no longer picks up the corresponding residue. Below we will interpret this transition as a kind of wall crossing phenomenon.

 \subsection{Phase diagram in the $(\mu, \beta)$-plane}

In figure 2 we have plotted the real part of the grand canonical partition function ${\cal Z}(\mu,\beta)$ as a function of the real-valued chemical potential $\mu$ and inverse temperature $\beta$,
\be\label{realvar}
{\rm Im} \,\rho  = {\mu \over 2\pi} \qquad \mbox{and} \qquad {\rm Im}\,\tau = { \beta \over 2\pi}
\ee 
 \begin{figure}[H]
		\includegraphics[width=10cm]{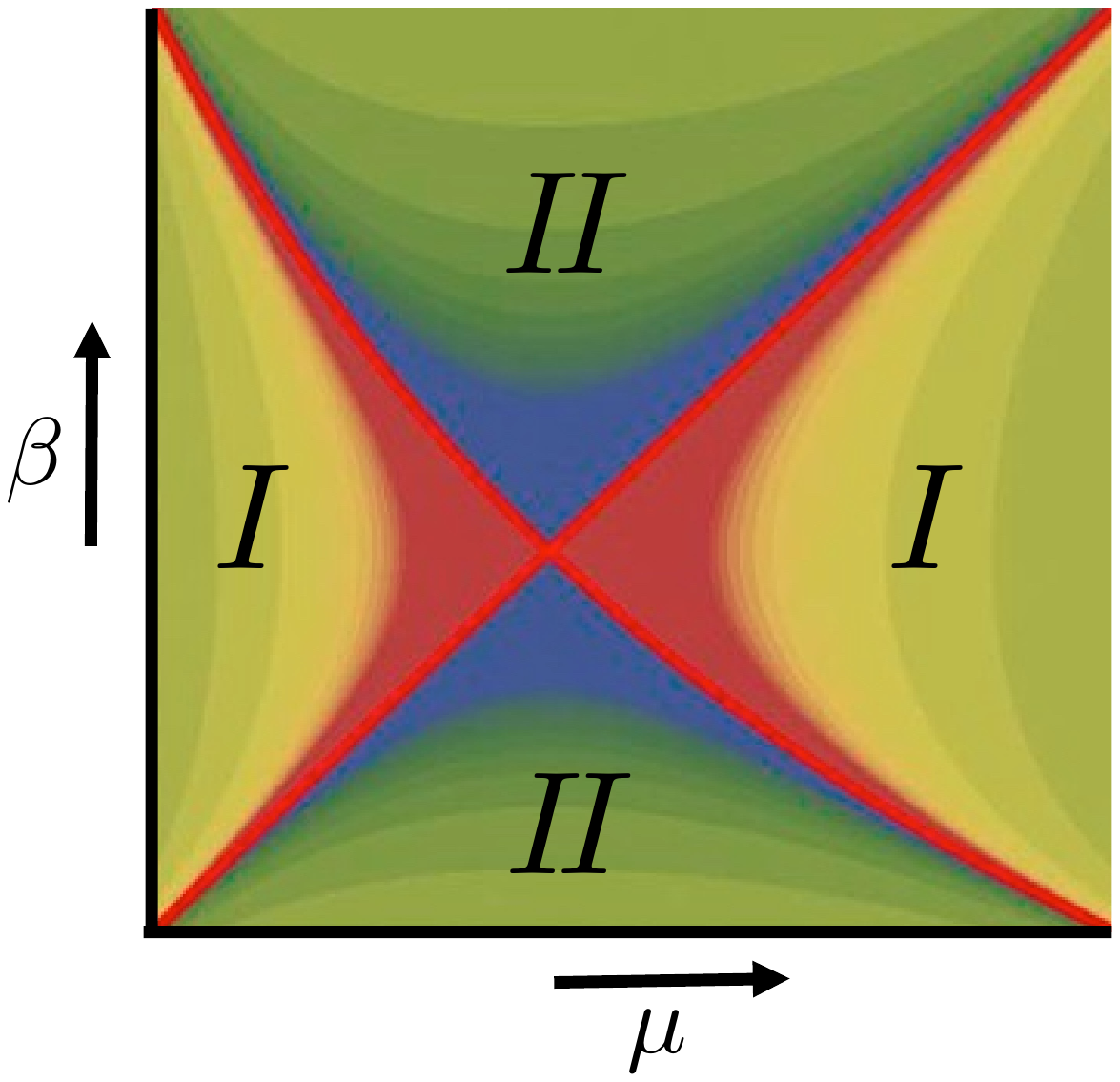}\vspace{-4mm}
		\centering
		\caption{Contour plot of the grand canonical partition function $\mathcal{Z}(\mu,\beta)$, where singularities of $\mathcal{Z}(\mu,\beta)$ are drawn in red. The partition function is positive in the regions labeled by $I$, and negative in the regions with label $I\!I$.
		}\label{fig:walls}
	\end{figure} 
This figure clearly exhibits the $\mathbb{Z}_2$ (anti-)symmetry that exchanges 
\be
  \mu \  \leftrightarrow\ \beta
\ee
which together with the invariance under the modular transformation of both $\mu$ and $\beta$
 \be
 \label{modular}
\mu \to {4\pi^2\over \mu} 
 \qquad\mbox{and}\qquad \beta\to {4\pi^2\over \beta} \ee 
generates a $D_4 = \left({\mathbb Z_2}\times {\mathbb Z_2}\right) \ltimes {\mathbb Z_2} $ subgroup of the full $O(2,2,\mathbb{Z})$ duality group. The red lines in the figure indicate the location of the wall at $\mu=\beta$ as well as its modular image $\mu=4\pi^2/\beta$. These walls separate the diagram in four chambers. The grand canonical partition function ${\cal Z}(\mu,\beta)$ is positive in the  regions labeled $I$ and is negative in chambers labelled $II$.   Each of these chambers still has a $\mathbb Z_2$ symmetry under the modular transformation of $\mu$ and $\beta$. Hence in total the figure contains eight fundamental domains that are related by the action of $D_4$. 
Only two of these fundamental domains are physical, in the sense that ${\mathcal Z}(\mu, \beta)$ correctly reproduces the known features of the symmetric products of the monster CFT.  This physical region lies in the chamber labeled $I$ below the diagonal, where the grand canonical partition function is positive and  $\mu$ and $\beta$ obey
\begin{equation}
\label{physical}
\mu> \beta > {4\pi^2\over \mu}.
\end{equation}
In this section we will investigate the interpretation of the other regions that are obtained by acting with $D_4$ on this physical region.

\newpage

\subsection{Microcanonical ensemble  with chemical potential} 

Traditionally, we interpret the grand canonical partition function as an expansion in powers of $p$ where each term is weighted by the canonical partition function $Z_N(q)$, as in ($\ref{Zdef}$). Equivalently, $Z_N(q)$ can be obtained by a contour integration over $p$.  This $p$-expansion and the contour prescription are only valid in the regime where $|p|<|q|$. Modular invariance of the canonical partition function implies that we should also require that $|p|<|\widetilde{q}|$. These two conditions define the physical region (\ref{physical}). 

In principle one can define a new  ensemble in which one fixes
the shifted conformal dimension $M=\Delta-c/24$ and the chemical potential $\mu$ dual to $N=c/24$, i.e. where one fixes the total energy while the total number of degrees of freedom is allowed to vary.  The partition function ${\widetilde{Z}}_M(p)$ for this ensemble is obtained from the grand canonical partition function ${\cal Z}(p,q)$ by  performing the contour integration over $q$  
\be
\label{tildeZdef}
{\widetilde{Z}}_M(p) = \oint\! {dq\over 2\pi i} \,q^{-M-1} {\cal Z}(p,q). 
\ee
It is important to point out that the contour in (\ref{tildeZdef}) has to be chosen so that it lies in the same physical region (\ref{physical}).  The reason for this is that we require that, by expanding ${\widetilde{Z}}_M(p)$ in powers of $p$, we obtain the same microscopic degeneracies $D(N,M)$ as given in (\ref{Zdef}), so that
\be
{\widetilde{Z}}_M(p)= \sum_{N\geq 0} D(N,M)\,p^{N+1}\\[-2mm]~.
\ee
To verify that this is indeed the case, let us compute the partition function ${\widetilde{Z}}_M(p) $ by explicitly performing the contour integral. For this purpose we again insert the expression (\ref{sumoverpoles}) for the grand canonical partition function, which can be written in the equivalent form
\begin{equation}
{\cal Z}(p,q) = \sum_{\gamma} {{\cal Z}_\infty (p_\gamma)
\over p_\gamma^{-1} - q^{-1}}\\[-2mm]
\end{equation}
The contour-integration over $q$ is straightforward.  The pole at $p=q$ does not contribute, since in the physical region $|p|<|q|$ the contour closes in the opposite direction, so 
\be
{\widetilde{Z}}_M(p) = -\sum_{\gamma\neq 1} p^{-M+1}_\gamma{\cal Z}_\infty (p_\gamma)\\[-2mm]
\ee 
Comparing this expression with the result (\ref{fareytail}) for the canonical partition function we find that the (anti-) symmetry between $p$ and $q$ is indeed broken since the contribution of the pole at $p=q$ is missing. By adding and subtracting this contribution we find that the two partition functions are related by
\be
\label{wallcrossing}
{\widetilde{Z}}_M(p) =p^{-M+1}{\cal Z}_\infty(p)- Z_{M-1}(p)
\ee
Here the first term represents the contribution of all the perturbative states. Furthermore, when $M<0$ the second term on the r.h.s. vanishes, hence in this regime the states are given purely by the perturbative spectrum. 

\subsection{A wall crossing formula}

The relation (\ref{wallcrossing}) can alternatively be written as as an expression for the canonical partition function
\be
\label{wallcrossing2}
Z_N(q) = q^{-N}{\cal Z}_\infty(q) - {\widetilde{Z}}_{N+1}(q)
\ee
We like to point out that all three partition functions have $q$-expansions with positive coefficients. 
By inserting the known expansions  (\ref{Zdef}) and (\ref{pertexp}) in to equation (\ref{wallcrossing}) or  (\ref{wallcrossing2}) 
one derives the following identity for the degeneracies $D(N,M)$ 
\be
\label{sumrule}
D(N,M)  = D_\infty(N+M)- D(M-1,N+1)~.
\ee
This relation generalizes the identity (\ref{negativeM-sumrule}) to general values of $M$ and implies that the degeneracies $D(N,M)$ are strictly less than (or in the case $M<0$ equal to) the asymptotic perturbative degeneracies $D_\infty(N+M)$.

The sum rule (\ref{sumrule}), as well as the relations (\ref{wallcrossing}) and (\ref{wallcrossing2}), show that the anti-symmetry between $p$ and $q$ of the grand canonical partition function translates at the micro-canonical level to an almost anti-symmetry under the exchange
\be
M\,  \leftrightarrow \, N+1 \label{MNflip}
\ee
This exchange symmetry is broken at the micro-canonical level, however, due to the contribution of the universal perturbative states $D_\infty(\Delta)$.

The sum rule (\ref{sumrule}) can also be verified directly from the expressions (\ref{DNM-sum}) and (\ref{Dinfty}) for $D(N,M)$ and $D(N+M)$ in terms of the degeneracies ${\cal D}(N,M)$ without the ground state contribution. 
One starts with the observation that the expression (\ref{Dinfty}) for the universal density can be written as a sum of two terms 
$$
D_{\infty} (N+M) =
\sum_{K=0}^{N} {\cal D}(N-K, M+K) +\sum_{K=1}^{M} {\cal D}(N+K, M-K)~.
$$
In the first term we recognize (\ref{DNM-sum}), while in the second term one can use the fact that ${\cal D}(N,M)$ is symmetric in its arguments to write the result in the same form as (\ref{DNM-sum}) with $M$ and $N$ shifted by 1. 

A third useful proof of (\ref{sumrule}) starts from the integral representation for the sum of the degeneracies 
$$
D(N,M) +D(M-1,N+1) = \oint {dp\over 2\pi i}\, \oint {dq\over 2\pi i}  \left({1\over p^{N+2}q^{M+1}} + {1\over p^{M+1}q^{N+2}} \right){\cal Z}(p,q)
$$
Here the contour is chosen so that the $p$ integral contour contains the contribution of the pole at  $p=q$ (as well as all other poles at $p=q_\gamma$). This can be achieved for instance by taking ${\rm Im}\rho > {\rm Im}\tau$. 
One subsequently uses the antisymmetry (\ref{antisym}) of the partition function  to rewrite the second integral on the right hand side by interchanging $p$ and $q$. At first it seems that the total result vanishes. However, while interchanging $p$ and $q$ one has displaced the integral contour so that it no longer contains the primary pole at $p=q$.  Hence the total result is given by the residue at the primary pole, which gives the universal perturbative degeneracy $D_\infty(N+M)$.

This third derivation makes clear that the difference between the degeneracies $D(N,M)$ and $D(M-1,N+1)$, and hence the broken exchange symmetry between $M$ and $N+1$, is due to the choice we made in equation (\ref{zerosum}) in expanding the partition function ${\cal Z}_0(p,q)$ of the ground states.  In the physical regime we imposed that $|p|<|q|$ and consequently the expansion is in positive powers of $p$. If instead we would have chosen $|q|<|p|$, the expansion would have been
\be
\label{zerosum2}
{\cal Z}_0(p,q) = 
-\sum_{K=0}^\infty q^{K+1} p^{-K} \qquad \mbox{for} \qquad |q|<|p| 
\ee
By exchanging the roles of $p$ and $q$ one has effectively removed the contribution of the $p=q$ pole from the canonical partition function. This situation is reminiscent of the wall crossing phenomenon, where certain marginally stable states disappear from the spectrum as moduli are varied
 \cite{Sen:2007vb, Cheng:2007ch}; see \cite{Cheng:2008fc} for a discussion of the phenomenon of wall crossing in the context of Borcherds algebras.

To explore this, let us define new degeneracies $\widetilde{D}(N,M)$ by expanding the grand canonical partition function in the unphysical region 
where $|p|>|q|$: 
\begin{equation}\label{wallZ}
{\cal Z}(p,q) = \mathop{\sum_{M\geq 1}}_{\hspace{1pt}N\geq -M} \widetilde{D}(N,M)p^{N+1}q^M\qquad \mbox{for} \qquad |q|<|p| ~.
\end{equation}
Note that the range of $N$ and $M$ are differen, since the ground state partition function is expanded as in (\ref{zerosum2}).
From the anti-symmetry ${\cal Z}(p,q)=-{\cal Z}(q,p)$ we find
\be
\label{Dtilde}	
\widetilde{D}(N,M) = - D(M-1,N+1) 
\ee
Hence, after wall crossing the new degeneracies are always negative. The fact that the arguments of the two degeneracies are related by the exchange symmetry (\ref{MNflip}) implies that also the Cardy and Hagedorn regimes interchanged: the first applies for $M<N$ while the latter is valid for $M>N$.  Finally, applying the sum rule (\ref{sumrule}) we find that the new and old degeneracies precisely differ by the perturbative Hagedorn states
\be
\label{jump}
D(N,M)-\widetilde{D}(N,M) = D_{\infty}(N+M)
\ee
These kind of relations are well known in the context of wall crossing with 
states of marginal stability.

Our results of this subsection restore the exchange symmetry $\mu \leftrightarrow\beta$ in the sense that the expansion of the grand canonical partition function is invariant under the exchange of $p$ and $q$ provided one also crosses the wall of marginal stability at $p=q$ and changes the sign of the coefficients. This combined operation leaves the expansion coefficients invariant, as is evident from (\ref{jump}).

\subsection{Restoring the $O(2,2;\mathbb{Z})$ duality symmetry.}

In this final subsection we will generalize the previous observation to the full $O(2,2;\mathbb{Z})$ duality group. First we will focus on the $D_4$ subgroup that acts on the  $(\mu,
\beta)$-plane. 
In Figure 3 we have depicted the phase diagram, where we included two orange lines  at $\mu=2\pi$ and $\beta =2\pi$ corresponding to the fixed points of  the two modular symmetry transformations of $\mu$ and $\beta$ given in (\ref{modular}).  
\begin{figure}[H]
		\includegraphics[width=10cm]{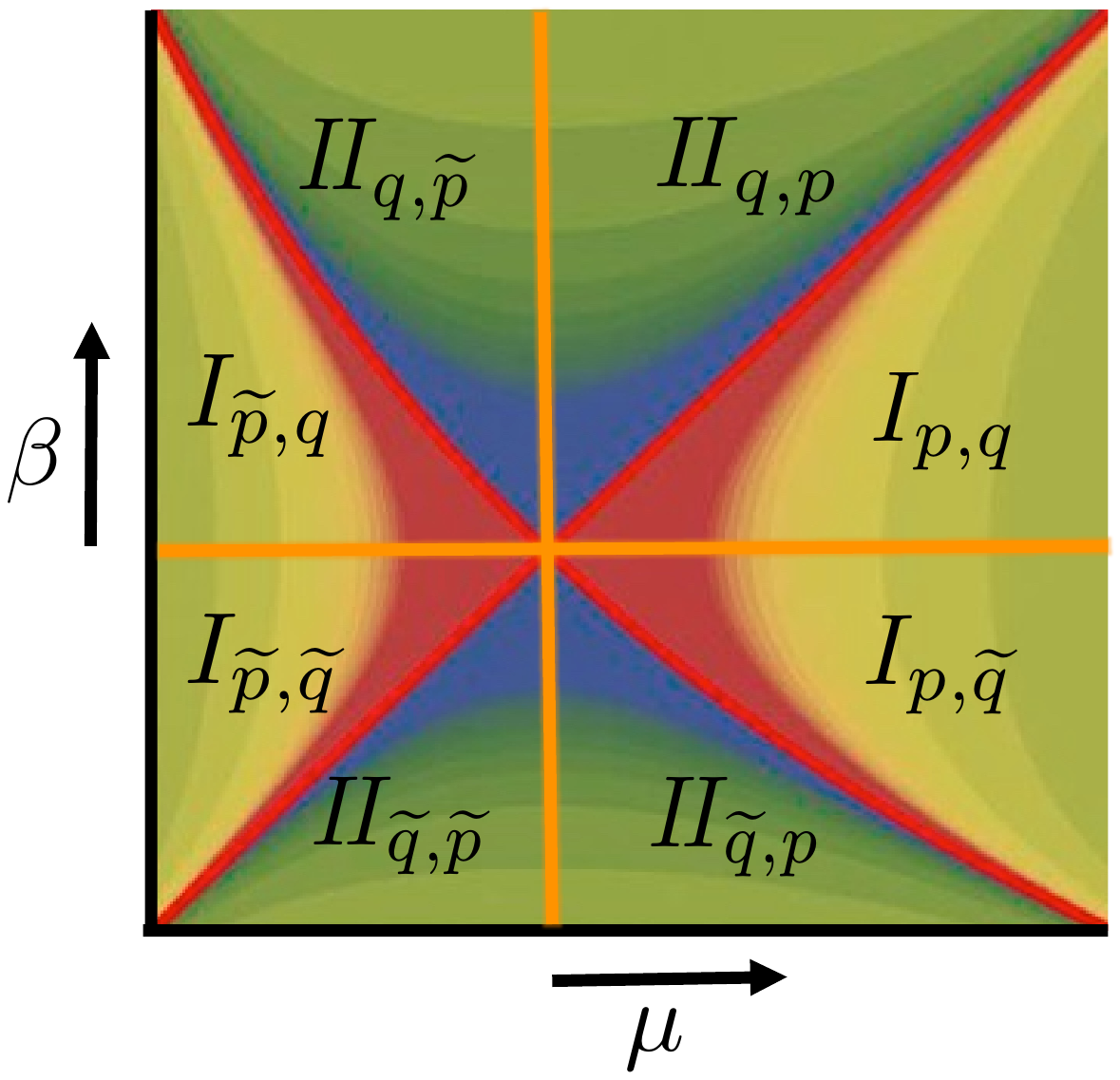}\vspace{-4mm}
		\centering
		\caption{The phase diagram of $\mathcal{Z}(\mu,\beta)$ with walls drawn in red and fixed lines of the modular transformations of $\mu$ and $\beta$ in orange. The labels in the 8 fundamental regions indicate the parameters in which the partition function needs to be subsequently expanded. 		}\label{fig:walls}
	\end{figure}  
 In addition to the wall at $\mu=\beta$, there is another wall at $\mu =4\pi^2/\beta$, or equivalently at $p=\tilde{q}$ where  $\tilde{q}=\exp(-4\pi^2/\beta)$. Similarly we will denote $\widetilde{p}=\exp(-4\pi^2/\mu)$. The orange and red lines divide the phase diagram in eight fundamental regions. The labels in each of these regions indicate in which variables the grand canonical partition function needs to be expanded.   For example, in the physical region indicated by $I_{p,q}$ the expansion is given first in terms of $p$ and subsequently in $q$. Note further that in this region $|q|<1$. The other regions and the corresponding expansion parameters are obtained by acting with the elements of the $D4$ symmetry group described in section 6.1. 

The discussion in the previous subsection dealt with the crossing of the $p=q$ wall from region $I_{p,q}$  in to the region labeled by  $I\!I_{q,p}$.  In this second region the partition function is first expanded in $q$ and then in $p$, since $|q|<|p|$.  As we explained, this leads to the expansion (\ref{wallZ}) in terms of the  coefficients $\widetilde{D}(N,M)$. The transition from region  $I_{p,q}$  to $I\!I_{q,p}$ is thus associated with a wall crossing phenomenon, in the sense that there is an integer jump in the value of the expansion coefficients.  As we will now explain, a similar phenomenon occurs when 
we cross the red line at $p=\widetilde{q}$. 

Let us consider the grand canonical partition function in the region labeled by $I_{p,\widetilde{q}}$ below the orange line. It is related to the original region $I_{p,q}$ by a modular transformation in $\beta$. The grand canonical partition function is invariant under this modular transformation, and hence the expansion in this region takes the form
\be
\mathcal{Z}(p,q)=\mathop{\sum_{N\geq 1}}_{\hspace{10pt}M\geq -N} D(N,M)p^{N+1}\widetilde{q}^M \qquad \mbox{for} \qquad |p|<|\widetilde{q}|<|\widetilde{p}|~.
\ee
Upon crossing the wall at $p=\widetilde{q}$, we enter into the region labeled by  $I\!I_{\widetilde{q},p}$. Hence, instead of exchanging $p$ and $q$ we have exchanged $p$ and $\widetilde{q}$.  This means that the expansion of the partition function in the region $I\!I_{\widetilde{q},p}$ is given by 
\be
\mathcal{Z}(p,q)=\mathop{\sum_{M\geq 1}}_{\hspace{10pt}N\geq -M} \widetilde{D}(N,M) p^{N+1}\widetilde{q}^M\qquad \mbox{for} \qquad |\widetilde{q}|<|p| <|q|
\ee
where the same relation (\ref{jump}) holds for the difference of the degeneracies. Hence, there is again a jump by the amount $D_\infty(N+M)$ in the expansion coefficients. 

An exactly analogous description of the expansions of the partition function holds in the other regions. 
We thus conclude that across the two walls in the $(\mu,\beta)$-plane, at $p=q$ and $p=\tilde{q}$, the expansion coefficients of $\mathcal{Z}(p,q)$ change precisely by the same amount set by the universal perturbative spectrum. The microscopic degeneracies take the same positive values $D(N,M)$ in all the regions in Figure 2 labeled by $I$ and the same negative values $\widetilde{D}(N,M)$ in all the regions labeled by $II$, where $\widetilde{D}(N,M)$ and $D(N,M)$ are related by (\ref{Dtilde}) or equivalently by (\ref{jump}). 

Note further that the walls at $p=q$ and $p=\tilde{q}$ are divided into two parts that related by the action of the modular group, which in the $(\mu,\beta)$-plane is given by $\mathbb{Z}_2$. But as we will now discuss, these observations are easily generalized to the full $O(2,2;\mathbb{Z})$-duality group that acts on the entire complex $(\rho,\tau)$-plane. In this plane the walls are located at all the $SL(2,\mathbb{Z})$ images of the $\rho=\tau$-wall, hence they are located at $\rho =\gamma(\tau)$. Alternatively, we can work in the complex space parametrized by $(p,q)$, in which some of the $O(2,2;\mathbb{Z})$ symmetries have already been modded out. In this complex $(p,q)$-plane the walls are located at $p=q_\gamma$. Each of these walls is divided in to fundamental domains labeled by a second $SL(2,\mathbb{Z})$-element. At these fundamental ``domain-walls" form the boundary between two fundamental regions of the full $(p,q)$ plane in which the partition function ${\cal Z}(p,q)$ is expanded in terms of $p_{\gamma_1}$ and $q_{\gamma_2}$ with $\gamma_1^{-1}\gamma_2 = \gamma$. On one side we have $|p_{\gamma_1}|<|q_{\gamma_2}|$ and the expansion is in powers of $p_{\gamma_1}$ and $q_{\gamma_2}$ with coefficients $D(N,M)$, while at the other side the expansion is in powers of the same variables but with coefficients $\widetilde{D}(N,M)$. In this way we have completely specified how the grand canonical partition function needs to be expanded to obtain a fully $O(2,2,\mathbb{Z})$ invariant description. 

\section{Discussion}

In this paper we discussed the family of two dimensional conformal field theories obtained by taking symmetric products of the $c=24$ chiral theory known as the `Monster CFT'. To do so we introduced the grand canonical ensemble of CFTs, with an extra chemical potential conjugate to the central charge $c=24N$, and demonstrated that this function is an automorphic function for $O(2,2;\mathbb{Z})$. We exploited this symmetry to give a rigorous derivation of the Cardy formula in the regime of large central charge $c$ and conformal dimension $\Delta>c/12$. In particular, we did not need to impose the usual restriction that the conformal dimension $\Delta$ is much larger than $c/24$, as is usually done,  in the proof of the Cardy formula.  

The family of conformal theories we considered have a sensible large $N$ limit, in the sense that they possess a well-defined perturbative spectrum of low-lying states in the large $c$ limit. This perturbative spectrum exhibits Hagedorn growth, while at high energies the spectrum is dominated by states with Cardy behavior. The Hagedorn growth seems to hint towards a higher spin or tensionless string realization of the putative gravitational dual theory.  We emphasize that the symmetric product theories we consider are not themselves extremal, and in fact are quite far from ``pure" theories of quantum gravity with only metric degrees of freedom. Indeed, it is not known whether extremal CFTs with large central charge exist (see e.g. \cite{Gaberdiel:2007ve, Gaberdiel:2008xb, Gaiotto:2008jt, Benjamin:2016aww}), and CFTs with large central charge are generically  expected to contain many higher spin primary operators with  conformal dimension $\Delta < c/24$.

The general set-up of the grand canonical ensemble in the gravitational picture would be one where we allow the cosmological constant $\Lambda$ to vary, or, more accurately, the dimensionless ratio $\sqrt{k}=R_{AdS}/\ell_s$ (where $\Lambda=-1/R_{AdS}^2$).  Such ensembles have been considered in a general context in \cite{Kastor:2009wy}.  In \cite{Giveon:2005mi}, a similar scenario was considered in the context of string theory in $AdS_3$. There it was shown that this theory has two phases: one phase that exhibits Cardy growth (the $k>1$ phase) and a phase where the long strings become weakly coupled and the spectrum develops Hagedorn behavior ($k<1$). It would be interesting to see if this transition can be understood microscopically by a generalization of our methods. Clearly, it is not expected that the particular ``Monster" conformal theories that we considered play a role in generic compactifications of string theory. But similar monstrous Lie algebras have been proposed to arise in certain CHL-compactifications of heterotic string theory, where they are realized as the algebra of BPS-states \cite{Paquette:2016xoo}.

It would be interesting to apply the presented results to a setting where we do have a clear holographic dual picture, preferably one that can be embedded in string theory. A natural candidate would be to start with the elliptic genus of strings on $K3$ and consider its grand canonical ensemble, and exploit its symmetries to derive the generalized Cardy regime. Important work along these lines was initiated in \cite{Benjamin:2015hsa} and \cite{Belin:2016knb}. 

In this paper we also discussed the physical meaning and implications of the apparent symmetry under the exchange of the inverse temperature and the chemical potential. As we described, this symmetry appears to be broken at the micro-canonical level. It is tempting to speculate, however, about a possible new phase that appears after wall crossing in which the roles of $p$ and $q$ and $N+1$ and $M$ are completely reversed. Such a transition is indeed present in the $D1$-$D5$ system that is used to count $N=4$ BPS black holes \cite{Dijkgraaf:1996it} via the 4D-5D connection \cite{Gaiotto:2005gf}. However,  unlike in that case the physical interpretation in our situation is somewhat more mysterious and troublesome. The new expansion (\ref{zerosum2}) of the ground state partition function has actually removed the ground states of the symmetric product CFTs and introduces new states with negative conformal dimension. Such CFTs are known to exist (for instance in the minimal models), but are generally expected to be non-unitary. Another puzzling aspect is that after wall crossing, due to the anti-symmetry of the grand canonical partition function, all the degeneracies are given by negative integers. Hence the  precise physical significance of the apparent exchange symmetry still remains somewhat obscure. \\[4mm]

\noindent\textbf{Acknowledgments:} We thank A. Belin, N. Benjamin, E. Dyer, A. Fitzpatrick, D. Kastor, C. Keller, P. Nayak, E. Perlmutter and J. Traschen for useful discussions.   
This research of AM is supported by National Science and Engineering Council of Canada and by the Simons Foundation, through the Simons Collaboration {\it It from Qubit}.

\appendix

\end{document}